\documentclass[final,notitlepage]{iopart}

\usepackage[dvips,colorlinks=true,pdfauthor={J. Brand}]{hyperref}
\usepackage{graphicx,bm,cite}

\newcommand{\be}{\begin{equation}}
\newcommand{\ee}{\end{equation}}
\newcommand{\bea}{\begin{eqnarray}}
\newcommand{\ba}{\begin{align}}
\newcommand{\eea}{\end{eqnarray}}
\newcommand{\ea}{\end{align}}
\newcommand{\bse}{\begin{subequations}}
\newcommand{\ese}{\end{subequations}}
\newcommand{\text}{\mathrm}

\begin{document}

\paper{A density-functional approach to fermionization in the 1D {B}ose
  gas}
\author{Joachim Brand}

\address{Max Plank Institut f\"ur Physik komplexer Systeme,
N\"othnitzer Stra{\ss}e 38, D-01187
Dresden, Germany}

\eads{\mailto{joachim@mpipks-dresden.mpg.de}}

%\submitto{\JPB {\bf 37} (2004) S287-S300 }
{\vspace{28pt plus 10pt minus 18pt}
     \noindent{\small\rm published in: {\it \JPB} {\bf 37} (2004)
     S287-S300\\misprints corrected September 2004\par}}

\date{\today}

\begin{abstract}
A time-dependent Kohn-Sham scheme for 1D bosons with contact
interaction is derived based on a model of spinor fermions.  This
model is specifically designed for the study of the strong interaction
regime close to the Tonks gas.  It allows us to treat the transition
from the strongly-interacting Tonks-Girardeau to the
weakly-interacting  quasicondensate regime and provides an
intuitive picture of the extent of fermionization in the system. An adiabatic
local-density approximation is devised for the study of time-dependent
processes. This scheme is shown to yield not only accurate
ground-state properties but also overall features of the elementary
excitation spectrum, which is described exactly in the Tonks-gas
limit.
\end{abstract}
\pacs{05.30.Jp,71.15.Mb,02.30.Ik,03.75.Kk}

\maketitle

\section{Introduction}

Density-functional theory (DFT) provides a unique framework for the
treatment of quantum many-body systems beyond the realm of
perturbation theory.  While ground-state DFT, based on the theorems of
Hohenberg, Kohn, and Sham \cite{hohenberg64,kohn65} deals with the
energy and the density profile of inhomogeneous ground states, the
time-dependent generalization of these theorems by Runge and Gross
\cite{runge84} in principle allows the study of time-dependent
processes and excited states. Further development of the
time-dependent DFT has shown that it is advantageous to consider
functionals simultaneously of the density and the current
\cite{vignale96} in order to go beyond the simplistic adiabatic
approximation and the microscopic Navier-Stokes equations for the
electron fluid have been derived elegantly this way \cite{vignale97}.

In this paper we will apply DFT to the study of fermionization in the
system of 1D bosons with contact interaction. In the case of infinite
interaction strength, the model is also known as the Tonks-Girardeau
gas \cite{girardeau60} and maps exactly to the system of
non-interacting spinless fermions with identical energy spectrum and
single-particle density. In fact, it has been shown by Cheon and
Shigehara \cite{cheon99} that the exact mapping between the original
Bose system and a Fermi system can be extended to arbitrary
interaction strength at the expense of a highly singular interaction
in the fermionic picture. For the homogeneous Bose gas, exact
solutions for stationary states at arbitrary interaction strength can be
found using the Bethe ansatz \cite{lieb63:1,lieb63:2,korepin93}.
Although the
exact solutions show a continuous transition between the perturbative
regime of weakly-interacting bosons and the strongly correlated,
fermionized regime, they do not prove very useful for the study of
time-dependent processes or inhomogeneous situations. Neither do they
provide us with a simple, intuitive picture of how strong the degree
of fermionization is in a given system.  The crossover from the
quasicondensate to the fermionized Tonks gas has also been studied in
Refs.~\cite{Olshanii1998a,petrov00}

Inspired by the works of Haldane \cite{haldane88}, Sutherland
\cite{sutherland71}, and others on the concept of exclusion statistics
describing a crossover between fermionic and bosonic statistics in 1D
systems, we 
study a model which explicitely allows this transition and provides an
intuitive picture of the degree of fermionization.  In order to devise
a practical scheme for treating time-dependent and inhomogeneous
systems we employ DFT and develop a time-dependent Kohn-Sham formalism
based on the auxiliary model system of $N$ non-interacting
spin-$(\nu-1)/2$ fermions. This model is chosen because the spin
degeneracy may simulate the level attraction or the bunching of
single-particle quasi momenta in the interacting
bosonic system \cite{lieb63:1}.
The interaction energy of the Bose system is simulated by the kinetic
energy of the spin-degenerate fermions.  In this work we will study
the model in the simplest and most generic approximation: the
adiabatic local density approximation (ALDA). The limiting case
of infinite interaction strength is obtained easily and is treated
exactly with $\nu=1$. The opposite limit of weak interaction can also
be treated accurately within the proposed formalism with
$\nu=N$ where the perturbative
Gross-Pitaevskii and Bogoliubov equations are recovered
asymptotically. In the general case of arbitrary interaction strength,
the spin degeneracy $\nu$ is fixed by requiring the correct low-energy
asymptotics of the excitation spectrum, which is analyzed in the
framework of linear-response theory. The resulting model is suited for
the study of time-dependent processes in inhomogeneous 1D Bose gases
close to the Tonks-gas limit. The properties of this approximate model
are analyzed and systematic improvements are suggested.

Before developing the general theory we want to point out that DFT is
not commonly used to discuss the theory of ultra-cold bosonic gases
although many standard approximations can be derived in this
framework. Only few papers therefore explicitely suggested to apply DFT
in this context, see
e.g.~Refs.~\cite{griffin95,nunes99,Kolomeisky2000a,oehberg02,kim03,chiofalo98,chiofalo01},
where the time-dependent DFT of superfluids
\cite{chiofalo98,chiofalo01} is particularly far developed. 
The case of strongly interacting bosons in 1D has been
considered by Kolomeisky {\it et al.}~\cite{Kolomeisky2000a} who
suggested a generalization of the 
Gross-Pitaevskii equation of the following form:
\begin{equation} \label{kmeq}
  i \hbar \frac{\partial}{\partial t} \psi(x,t) = \left\{ -
    \frac{\hbar^2}{2 m} \frac{\partial^2}{\partial x^2} +
     v_{\text{ext}}(x,t) + \phi(n(x,t))
    \right\} \psi(x,t)
\end{equation} 
Here, $\psi(x,t)$ is a time-dependent complex field and $n(x,t) =
|\psi(x,t)|^2$ is the one-dimensional density. The function
$\phi(n)$
provides the nonlinear term. 
Kolomeisky {\it et al.} suggested to use the chemical potential of
the Tonks-Girardeau gas  
\be
  \phi(n) = \phi^{{\mathrm (TG)}}(n) \equiv \frac{\pi^2 \hbar^2}{2 m} n^2 
\ee 
whereas the Gross-Pitaevskii equation
is recovered for
\be
  \phi(n) = \phi^{{\mathrm (GP)}}(n) \equiv g_{{\mathrm 1D}} n .
\ee 

We will refer to Eq.~(\ref{kmeq}) as the {\em bosonic LDA} because this
equation may be derived as a Kohn-Sham equation in the ALDA using a
Bose condensate as the auxiliary non-interacting system for the Tonks
gas and the weakly interacting Bose gas, respectively.  The same
equation also applies for 1D bosons with arbitrary interaction
strengths and was used by {\"O}hberg and Santos \cite{oehberg02}. In
this case, $\phi(n)$ represents the exactly known chemical potential of
the Lieb-Liniger model (at $T=0$). The bosonic LDA equation
(\ref{kmeq}) is almost identical to a set of hydrodynamic
equations\cite{dunjko01,menotti02a} derived from general hydrodynamic
arguments assuming local equilibrium, the only difference being a
kinetic energy pressure term (see e.g.~\cite{fetter01}).

It may be questioned whether the
choice of a Bose condensate as the auxiliary non-interacting system of
the Kohn-Sham formalism is the ideal starting point for treating a
strongly interacting Bose gas in 1D which is not
condensed\cite{lenard64,girardeau60}.  Indeed it has been shown in the
Tonks-Girardeau limit of impenetrable point bosons, that phase
coherence is grossly overestimated by 
Eq.~(\ref{kmeq}) leading to wrong predictions for interference
properties \cite{girardeau00a,damski03ep:1}. Phase and density are
conjugate variables in a quantum field theory and may coexist in the
weakly-interacting Bose condensate, which is close to a
classical field.  It is known already in the perturbative regime, that
phase fluctuations are very important in 1D \cite{popov83,petrov00}. In the
ultimate limit of the Tonks-Girardeau gas, the concept of phase looses
its meaning and the governing equations should be rather based on the
density alone. For this reason the hydrodynamic approximations
employed in Refs.~\cite{dunjko01,menotti02a}, where the concept of a
phase does not appear, are on a safer footing than the bosonic LDA
equation of Kolomeisky where phase-coherence is built into the
theory.

Moreover, the excitation
spectrum and density of the the Tonks-Girardeau gas is identical to
the corresponding properties of a gas of independent spinless fermions
due to the Bose-Fermi mapping.
Already in his original paper\cite{lieb63:1}, Lieb pointed out, that the
fermionic character of the excitation spectrum prevails at finite
interaction strength and introduced two elementary branches of
excitations which he called type~I and type~II. Type I excitations
result from exciting a particle from the edge of the 1D Fermi sphere
to an unoccupied orbital. Thus type I excitations are the
particle-hole excitations
of the highest possible energy for a given momentum. Type~II
excitations, on the contrary, are the excitations of the lowest
possible energy for a given momentum. They can be seen as taking a
particle from the inside of the Fermi sphere and putting it to the
first free orbital on the outside.

Only type~I excitations are found by standard bosonic methods like
Bogoliubov perturbation theory or linear response in Eq.~(\ref{kmeq})
whereas type~II excitations and the fermionic character of the
spectrum remain hidden. We want to mention at this point that a
possible connection between type~II excitations and dark solitons in
variations of Eq.~(\ref{kmeq}) has been pointed out in the literature
\cite{kulish76,Kolomeisky2000a,jackson02,komineas02}. 
%The exact dynamics of
%dark solitons in the strongly-interacting 1D Bose gas has been further
%discussed in references \cite{Girardeau2000b,girardeau03,busch03}.

\section{\label{sec:exotic} An exotic Kohn-Sham scheme}
\subsection{Variational principles}
The basic scheme of DFT is the Hohenberg-Kohn variational principle.
It allows the ground-state energy $E_0$ and one-particle density
$n(x)$ of a given many-particle system to be found by variationally
minimizing the energy functional
\begin{equation} \label{HK}
  E_v[n] = F_{\text{HK}}[n] + \int v_{\text{ext}}(x) n(x) dx .
\end{equation}
Here $v_{\text{ext}}(x)$ is the external potential and the
Hohenberg-Kohn functional
$F_{\text{HK}}[n]$ is universal in that it does not depend on the
external potential. 
The scheme of Kohn and Sham proceeds by partitioning the unknown
functional $F_{\text{HK}}[n]$  as
\begin{equation}
  F_{\text{HK}}[n] \equiv T_s[n] + E_{\text{KS}}[n],
\end{equation}
where $T_s[n]$ is the kinetic energy of a ficticious, auxiliary
system of
particles interacting only with a single-particle potential $v_s$,
which has the same density $n(x)$ as the original system. 
The external potential
$v_{\text{ext}}$ is contained in 
$v_s = v_{\text{ext}} + v_{\text{KS}}$ where $v_{\text{KS}}[n] = 
\delta  E_{\text{KS}} / \delta n$ is a
universal functional of the density 
depending only on the structure and internal interactions of the
original and auxiliary many-body problem but not on the external
potential. Once an (approximate) expression for $v_{\text{KS}}[n]$
has been found, the density of the system in a given external
potential $v_{\text{ext}}$ is easily found by solving the Kohn-Sham
equations, a set of nonlinear independent-particle equations.
Kohn and Sham in their original publication\cite{kohn65} used the
auxiliary system of noninteracting spin-1/2 fermions while interested
in systems of interacting electrons. However, it has been pointed out
that the choice of the auxiliary system is somewhat arbitrary and
rather a matter of convenience than dictated by laws of nature
\cite{levy84}. 

\subsection{Hamiltonian}

We start with the usual Hamiltonian for $N$ bosons in 1D with point
interactions of Ref.~\cite{lieb63:1}, augmented with a possibly
time-dependent external potential $v_{\text{ext}}(x,t)$:
\begin{equation} \label{llham}
  H=  \sum_{i=1}^N \left\{ -\frac{\hbar^2}{2 m} \frac{\partial}{\partial x_i}
      + v_{\text{ext}}(x_i,t) \right\} + 2 g_{\text{1D}} \sum_{i
      < j} \delta(x_i - x_j) ,
\end{equation}
where $g_{\text{1D}} = 2\hbar^2/(m |a_{\text{1D}}|)$ is the interaction
parameter and
$a_{\text{1D}}$ is
the one-dimensional scattering length\cite{Olshanii1998a}. 

The exact many-body eigenstates and the complete excitation spectrum
of the Hamiltonian~(\ref{llham}) in the homogeneous system where
$v_{\text{ext}}=0$ with periodic boundary conditions have been found
by Lieb and Liniger using the Bethe ansatz \cite{lieb63:1,lieb63:2}.
The ground state energy is 
\be
E_0^{\text{hom}} = N \frac{\hbar^2 n^2}{2 m}
e(\gamma)
\ee
where $e(\gamma)$ is
the dimensionless energy per particle in the Lieb-Liniger model and
$\gamma = 2/(n 
|a_{\text{1D}}|)$ is the single dimensionless parameter of the
homogeneous gas with the one-dimensional single-particle density $n =
N/L$ of $N$ particles in a 
box of length $L$.  The function $e(\gamma)$ is
defined as the solution of a Fredholm equation and can be obtained
numerically to any desired accuracy.
The chemical potential of the homogeneous Lieb-Liniger gas at density
$n$ is given by
\be \label{eqn:mu}
  \mu_{\text{LL}}(n) = \frac{d}{d N} E_0^{\text{hom}} =
  \frac{2 \hbar^2}{m  |a_{\text{1D}}|^2} f(\gamma), 
\ee
where $f(\gamma) = [3 e(\gamma) - e'(\gamma)]/\gamma^2$ is a dimensionless
function. The asymptotic behaviour of $e(\gamma)$ and $f(\gamma)$ is known
for large and small $\gamma$ \cite{lieb63:1} and, furthermore, the
functions are tabulated for intermediate values of $\gamma$ in
Ref.~\cite{dunjko01}
\footnote{The dimensionless chemical potential $f(\gamma)$ is
known to have the expansion\cite{lieb63:1} $f(\gamma) = \pi^2 \gamma^{-2} -
\frac{16 \pi^2}{3} \gamma^{-3} + 
{\mathcal{O}}(\gamma^{-4})$ for $\gamma \to \infty$ and $f(\gamma)
= 2 \gamma^{-1} - \frac{2}{\pi} \gamma^{-\frac{1}{2}} + 
{\mathcal{O}}(\gamma^{1})$ for $\gamma \to 0$.
We give the following convenient rational approximation $f(\gamma) \approx
[{\pi^2 (48 + 2 \gamma^2 + 3 \gamma^3)} ] / 
\{3 \gamma [8 \pi^2 + 8 \pi \sqrt\gamma + \gamma (2 + \gamma)^3]\}$,
which deviates with
a maximum relative error of $0.12$ from the numerical result at
$\gamma \approx 2$. It has the same expansion as given above for
$\gamma \to \infty$. For small $\gamma$ it introduces an error of
order ${\mathcal{O}}(\gamma^{1/2})$.
}.

\subsection{Functionals and Kohn-Sham equations}

Consider a physical system of $N$ noninteracting and indistinguishable
fermions with spin 
$(\nu - 1)/2$. For simplicity let us consider the case where $\nu$
divides $N$ and assume that all orbitals are nondegenerate.  In the
ground state $\tilde{N}$ orbitals will be occupied with $\nu$
particles each when $\nu \tilde{N} = N$. The kinetic energy is
\[
 T_s[n] = -\frac{\hbar^2}{2 m}\nu \sum_{i=1}^{\tilde{N}} \int
   \varphi_i^*(x) \frac{\partial^2}{\partial x^2} \varphi_i(x)\; dx .
\]
The Kohn-Sham equations are found by functional differentiation with
respect to $\varphi_i^*(x)$:
\begin{equation}\label{KohnShamIndep}
 \varepsilon_i \varphi_i(x) = \left\{ -\frac{\hbar^2}{2
m}\frac{\partial^2}{\partial x^2} + v_{\text{ext}}(x) +
 v_{\text{KS}}[n](x) \right\} 
 \varphi_i(x) . 
\end{equation}
The potential $v_{\text{KS}}[n] = \frac{\partial
E_{\text{KS}}[n]}{\partial n}$ is the mean-field potential of
Kohn-Sham theory and represents the unknown rest of the functional. In
regular electronic DFT, $v_{\text{KS}}[n]$ would be written as a sum
of the Hartree potential and the exchange-correlation potential. The
density is found self-consistently by summing over all occupied
($\in$) orbitals
\[
  n(x,t) = \nu \sum_{i \in} |\varphi_i(x,t)|^2 
\]
The ground state density $n(x)$ is obviously time-independent and only the
$\tilde{N}$ levels with the lowest energies $\varepsilon_i$ will be
occupied. The generalization to the corresponding time-dependent
Kohn-Sham equations along the lines of the Runge-Gross theorems
\cite{runge84} is obvious: 
\begin{equation}\label{KohnSham}
 i \hbar \partial_t \varphi_i(x,t) = \left\{ -\frac{\hbar^2}{2
m}\frac{\partial^2}{\partial x^2} + v_{\text{ext}}(x,t) +
 v_{\text{KS}}[n](x,t) \right\} 
 \varphi_i(x,t) . 
\end{equation}
We note that
Eq.~(\ref{KohnSham}) is still exact. We now proceed in approximating
$v_{\text{KS}}[n]$ in a local-density approximation. 

To this end, the remaining unknown term $E_{\text{KS}}[n]$ is
approximated by $E_{\text{KS}}[n]\approx
E_{\text{KS}}^{\text{LDA}}(n) = E_0^{\text{hom}} - T_s^{\text{hom}}$,
where $T_s^{\text{hom}}$ is the noninteracting kinetic energy of the
homogeneous gas.
It is simply found by summing the
contributions from the $\tilde{N}$ orbitals in the Fermi sphere and
multiplying with the degeneracy $\nu$:
\begin{eqnarray} \label{Ekin}
\nonumber
 T_s^{\text{hom}} &=& \nu  \frac{\hbar^2}{2 m}
\sum_{k=-(\tilde{N}-1)/2}^{(\tilde{N}-1)/2} \left(\frac{2 \pi
  k}{L}\right)^2 \\\label{Ekin2} 
 & = & N  \frac{\hbar^2 \pi^2 {n}^2}{6 m \nu^2} \left(1 -
\frac{1}{\tilde{N}^2}\right) 
\end{eqnarray}
where $L$ is the length of a box with periodic boundary conditions, $n
= {N}/L = n$ is the line density and $\tilde{N}$ has been assumed odd.

Before taking the derivative $\frac{\partial
F_{\text{HK}}[n]}{\partial n}$ in order to find an expression for
$v_{\text{KS}}$ we have to consider what to do with the term
$1/\tilde{N}$. If $\tilde{N}=1$ the contribution from the kinetic
energy vanishes and we arrive at the bosonic LDA of
Eq.~(\ref{kmeq}). However, if $\tilde{N}$ is of the order of $N$,
we may disregard the term $1/\tilde{N}$ in the limit of large $N$. In
fact it is not consistent to keep this term if we approximate the
Lieb-Liniger energy with the bulk value as this is only valid in the
limit $N\to\infty$. In the following we will therfore drop this term
and arrive at
\begin{eqnarray} \label{eqn:vhxc}
  v_{\text{KS}}^{\text{LDA}}(x,t) &=& \frac{d}{d N} \left(E_0^{\text{hom}}
         - E_{\text{kin}} \right)|_{n =
         n(x,t)}\nonumber \\ 
    &=&  \mu_{\text{LL}}(n) - \frac{\hbar^2 \pi^2 n^2}{2 m \nu^2} ,
\end{eqnarray}
where the chemical potential $\mu_{\text{LL}}(n)$ of the homogeneous
(Lieb-Liniger) gas at density $n$ is given by Eq.~(\ref{eqn:mu}).  The
remaining parameter $\nu$ will be determined by a consistency condition
on the excitation spectrum that will be derived in the following
paragraph.

\subsection{Excitation spectrum from linear-response theory}
\label{sec:linearresponse}

Lieb pointed out that the excitation spectrum of the 1D Bose
gas resembles that of the 1D Fermi gas and likewise has a
particle-hole type of structure --- at any value of the interaction
strength \cite{lieb63:2}! The spectrum of the homogeneous gas may be
equally well 
described by two branches of elementary excitations
$\epsilon^{{\mathrm I}}(k)$ and $\epsilon^{{\mathrm{II}}}(k)$. Both
branches have the {\em same slope} for $k\to 0$ and therefore lead to the
{\em same speed of sound} $v_s = d \epsilon^{{\mathrm I}}/d k|_{k=0} = d
\epsilon^{{\mathrm{II}}}/d k|_{k=0}$. Bogoliubov perturbation theory
yields an approximation only for the branch $\epsilon^{{\mathrm
I}}(k)$.  In the time-dependent Kohn-Sham model introduced above, a
particle-hole type excitation spectrum arises naturally. The
elementary excitations may be found by solving for the resonant
frequencies of the time-dependent equation (\ref{KohnSham}) in the
linear-response 
limit. The resulting excitation energies are given by differences of
the orbital energies of the stationary Kohn-Sham orbitals modified by
interaction contributions from the nonlinear mean-field
$v_{\text{KS}}^{\text{LDA}}$.

\begin{figure}
\includegraphics[width=10cm]{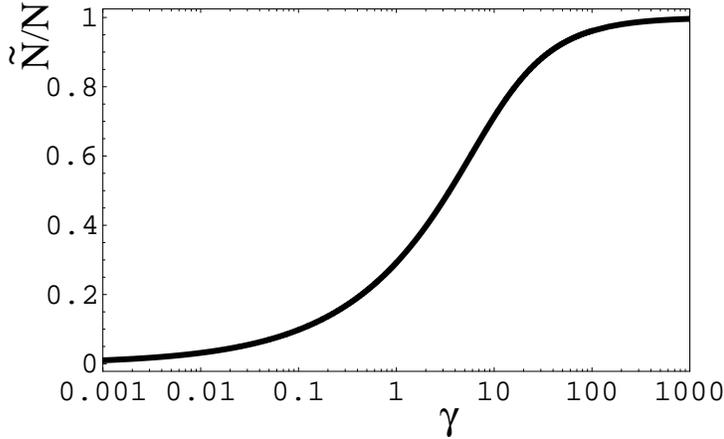}
\caption{\label{f:crossover} The parameter $\tilde{N}/N$ from
Eq.~(\ref{eqn:nu_opt})  indicates the
degree of fermionization in the system. Here it is plotted as a
function of the dimensionless interaction strength of the Lieb-Liniger
model $\gamma$.}
\end{figure}

The derivation of the linear response equations is the same as for the
LDA in the much studied electronic systems and very similar to the
time-dependent Hartree-Fock or random-phase approximation (see
e.g. \cite{thouless72}) and is summarized in the appendix. Here we
want to note that the linear response correction to the spectrum of
the Kohn-Sham system is proportional to the derivative
$dv_{\text{KS}}/d n$. From the explicit linear response equations for
the homogeneous system it can be seen that type I and type II
excitations have a different slope at $k=0$, except for the
situation where 
\be \label{eqn:noLR}
  v_{\text{KS}}'(n)=0 .
\ee 
This argument is detailed in the Appendix where the linear response
equations in the small momentum limit are solved exactly.
Using the expression (\ref{eqn:vhxc}) for $v_{\text{KS}}$
we easily find the correct value for the
number of modes in the fermionic Kohn-Sham equations (\ref{KohnSham}):
\be \label{eqn:nu_opt} 
  \tilde{N} = N \sqrt{- \frac{\gamma^3}{2 \pi^2} f'(\gamma) } .
\ee 
The value of $\tilde{N}/N \equiv
1/\nu$ is a measure of the degree of fermionization in the
system. When $\tilde{N}/N =1$ the system is completely fermionized whereas
$\tilde{N}/N \to 0$ indicates a purely bosonic system with a Bose condensate
as the ground state. The functional dependence of $\tilde{N}/N$ on the
interaction strength $\gamma$ is shown in Figure
\ref{f:crossover}. 

Qualitatively the picture agrees with the results of Girardeau and
Wright \cite{girardeau01a} who modelled the BEC-Tonks crossover by a
mixture of a pair-correlated Bose liquid with a non-interacting Fermi
gas in a variational framework.  We would like to point out, that
$\tilde{N}$ is a global parameter whereas $\gamma$ is local and varies
in an inhomogeneous system. Therefore the condition (\ref{eqn:nu_opt})
should be used with care and we expect the present model to be most
useful for nearly homogeneous systems.

Although it would be desirable to find an approximation for the
condensate fraction from our model, it should be noted that Kohn-Sham
DFT as used here generally does not model the many-body wavefunction
but gives direct access only to energies and the diagonal part of the
single-particle density matrix. The information about off-diagonal
long range order and condensate fraction, however, is contained in the
inaccessible off-diagonal part of the single-particle density
matrix. Nevertheless it should be pointed out that the fractional
occupation of the Kohn-Sham orbitals $\tilde{N}^{-1}$ in our model,
which is reminiscent of a condensate fraction, vanishes in the
thermodynamic limit where $N \to \infty$ at $\gamma = const.$
consistent with the general result that there is no Bose condensate in the
$T=0$ homogeneous interacting Bose gas \cite{schwartz77,haldane81}.

An important question related to the condensate fraction is local
phase coherence \cite{girardeau00a,damski03ep:1}. From the
construction of the Kohn-Sham model it appears difficult to define
expectation values for the phase operator in this scheme. However, the
time-dependent DFT does contain information about local phase
coherence to the degree that is necessary to correctly describe the
supression of interference patterns in the dynamics.

The approximation scheme defined by Eqs.~(\ref{KohnSham}),
(\ref{eqn:vhxc}), and (\ref{eqn:nu_opt}) is the main subject of this
paper and shall be referred to as {\em fermionic LDA}. In the
remainder of this paper we will analyze the predictions of this model
and compare with the exact solutions of the Lieb-Liniger model and
other approximations.

\section{\label{sec:props}Properties}
\subsection{Excitations of the homogeneous gas}
It is interesting to calculate the dispersion for the elementary
excitations of type I and II, which can be done analytically since
the linear-response corrections vanish due to Eq.~(\ref{eqn:noLR}).
In the homogeneous gas, the mean field $v_{\text{KS}}$ is constant and
the Kohn-Sham single-particle energies 
become
\[
  \epsilon_p = \frac{2 \pi^2 \hbar^2}{m L^2} p^2 + v_{\text{KS}} ,
\]
where $p$ is any integer and $\hbar k_p = \hbar p\, 2\pi/L$ is the
momentum. In the ground state all orbitals with $p < F$ are occupied
where $F=(\tilde{N}-1)/2$ is the index of the Fermi momentum $\hbar
k_F$. The energies 
of particle-hole excitations of momentum $\hbar k_q$ are given by the
difference
\begin{eqnarray} \label{eqn:phexcitations}
  \epsilon_p -\epsilon_{p-q} &=&\frac{\hbar^2}{2 m} n^2 \frac{4
  \pi^2}{N^2}  q (2p - q) . 
\end{eqnarray}

\begin{figure}
\includegraphics[width=10cm]{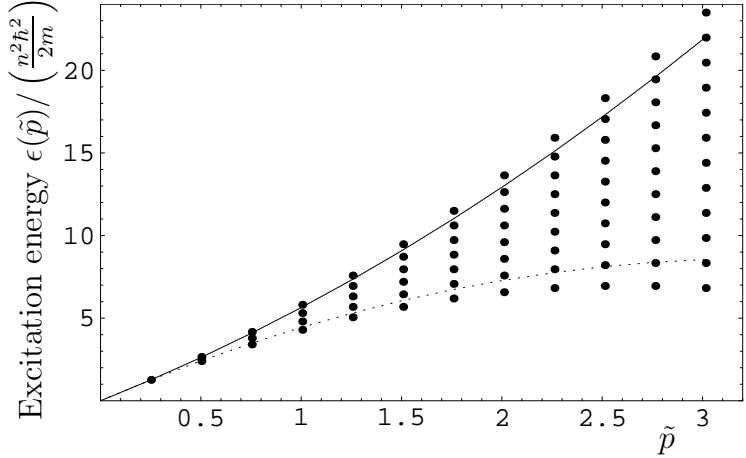}
\caption{\label{f:spectrum} Shown are excitation energies of a
homogeneous system with 25 particles at $\gamma = 16.3$. Dots indicate
the particle-hole excitation in the Kohn-Sham model with $\tilde{N}/N
= 0.8$ from Eq.~(\ref{eqn:phexcitations}). The full and dotted lines
show the exact type I and II 
excitations for this
finite system from the numerical solution of Yang's equation
\cite{lieb63:2,yang69}, respectively.}
\end{figure}

Figure \ref{f:spectrum} illustrates the discrete excitation spectrum obtained
for a finite number of particles in a box with periodic boundary
conditions. Type~I and II excitations are the upper and
lower bounds of the elementary excitation spectrum, respectively.
The fermionic DFT is seen to
slightly overestimate the energies of type I excitations and underestimate
the energies of type II excitations at large momentum while the
correct asymptotes are obtained for small momenta.  

From Eq.~(\ref{eqn:phexcitations}) we can construct explicit
expressions for  type~I and type~II
excitations. 
Type I excitations are defined by exciting a particle
with the Fermi momentum $\hbar k_F$ into an unoccupied orbital with
$p=F+q$ whereas 
type~II excitations take a particle from inside the Fermi sphere to the
lowest unoccupied orbital with $p= F+1$. 
%For type~I we have $p=F+q$, where $F=(\tilde{N}-1)/2$
%is the index of the Fermi wavenumber. 
In terms
of the dimensionless momentum
\[
  \tilde{p} = \frac{\hbar k_q}{\hbar n} = \frac{2 \pi q}{N}
\]
of the excitations we obtain
\begin{equation} \label{epsI}
  \epsilon^{\text{I}}(\tilde{p}) = \frac{\hbar^2}{2m} n^2 \tilde{p} 
    \left[  2 \pi \left(\frac{1}{\nu} -
    \frac{1}{N}\right) + \tilde{p}\right] .
\end{equation}
Correspondingly we find for type~II
\begin{equation} \label{epsII}
  \epsilon^{\text{II}}(\tilde{p}) = \frac{\hbar^2}{2m} n^2 \tilde{p} 
    \left[  2 \pi \left(\frac{1}{\nu} +
    \frac{1}{N}\right) - \tilde{p}\right] .
\end{equation}
%According to Lieb's
%classification \cite{lieb63:2}, elementary excitations of type~II
%are restricted to values of $q<F$ or $\tilde{p}<\pi$.

Expressions for the
elementary excitation spectrum of the interacting gas can be derived
using equation (\ref{eqn:nu_opt}). In the
thermodynamic limit we obtain
\begin{eqnarray} \label{eqn:TI}
   \epsilon^{\text{I}}(\tilde{p}) &=& \frac{\hbar^2}{2m} n^2 \tilde{p} 
     \left[ \sqrt{- 2 \gamma^3 f'(\gamma)}  + \tilde{p}\right] ,\\
\label{eqn:TII}
   \epsilon^{\text{II}}(\tilde{p}) &=& \frac{\hbar^2}{2m} n^2 \tilde{p} 
     \left[ \sqrt{- 2 \gamma^3 f'(\gamma)}  - \tilde{p}\right] .
 \end{eqnarray}
These dispersion relations ought to be compared with the exact ones
and with earlier approximations.
First of all we note that both
dispersions relations indeed lead to the same speed of sound
\be \label{eqn:sound}
  \left.\frac{d \epsilon^{\text{I/II}}}{dp}\right|_{p=0}  = c_s \equiv
   \frac{2\hbar}{m 
  |a_{\text{1D}}|} \sqrt{-\frac{\gamma}{2} f'(\gamma)} ,
\ee
which is identical to the exact result of the Lieb-Liniger
model\cite{lieb63:2}.

The type II excitations, as discussed by
Lieb, have the character of elementary excitations for $\tilde{p} \le
\pi$ and at this point the slope $d \epsilon^{\text{II}} /
d\tilde{p}$ vanishes. This condition is strictly fulfilled by
Eq.~(\ref{eqn:TII}) only in the Tonks-gas limit $\gamma\to\infty$
where the equations become exact. In the general case,
$\epsilon^{\text{II}}(\tilde{p})$ of Eq.~(\ref{eqn:TII}) has a maximum
at $\tilde{p}_m < \pi$ and $\tilde{p}_m$ approaches 0 as $\gamma\to
0$. This observation is related to the fact that the umklapp
transition of taking one particle from one side of the Fermi-sphere to
the other is not described correctly in the considered fermionic
Kohn-Sham model. Therefore our approach is limited to describing type
II excitations in the long-wavelength limit and works best for large
$\gamma$. 

\subsection{Comparison with bosonic LDA}

The bosonic LDA of Eq.~(\ref{kmeq}) is simply the single-mode case 
with $\tilde{N}=1$ of the general fermionic Kohn-Sham formalism in
this paper.
The linear-response equations (\ref{EinsteinBog}) for this case coincide
with the Bogoliubov-de Gennes equations derived in
Refs.~\cite{oehberg02,kim03}. It is instructive to derive explicit
expressions for the excitations of a homogeneous
gas, which appear not to be available in the
literature. Eqs.~(\ref{EinsteinBog}) now become 
% \begin{subequations} 
 \begin{eqnarray} \label{typeI}
   (\epsilon_q - \epsilon_0 + \frac{N \phi'}{L})u_q + \frac{\nu
     \phi'}{L} v_q &=& \hbar \omega  u_q \\ \label{typeII}
   (\epsilon_q  - \epsilon_0 + \frac{N \phi'}{L})v_q + \frac{\nu
     \phi'}{L} u_q &=& - \hbar \omega  v_q 
 \end{eqnarray}
%\end{subequations}
where $\phi'=d\phi/dn$ and we have set $X_{hp} = u_{p-h}$ and  $Y_{hp}
= v_{h-p}$.  With 
\[
  \epsilon_q  - \epsilon_0 = \frac{2 \pi^2 \hbar^2 q^2}{m L^2}
\]
we find the simple solution
\begin{equation} \label{elemexPhys}
  \hbar \omega = \pm\sqrt{\frac{\hbar^2 k_i^2}{2 m} (\frac{\hbar^2
    k_i^2}{2 m} + 2 n \phi')} .
\end{equation}
Only the plus sign contributes here, the minus sign is a well
understood artifact of linear response theory. 
In terms of the dimensionless momentum $\tilde{p} = k / n$ we
find the following final result for the excitation spectrum of the
bosonic LDA with $\phi(n) = \mu_{\text{LL}}(n)$:
\begin{equation} \label{elemex}
  \hbar \omega =  \frac{\hbar^2}{2 m} n^2 \sqrt{\tilde{p}^2
     (\tilde{p}^2 - 2 \gamma^3 f'(\gamma))}.
\end{equation}
In contrast to
the excitation spectrum described by Lieb, there is only one
excitation branch in the linear response of the bosonic LDA.
Comparing with Eqs.~(\ref{typeI}) and (\ref{typeII}) we see that the
first terms in the expansions around $k=0$ and $k=\infty$ are
identical but there are discrepancies in between.
The speed of sound is again the exact result $c_s$ as a consequence of
the compressibility sum rule which is obeyed by the LDA.

\begin{figure}
\includegraphics[width=10cm]{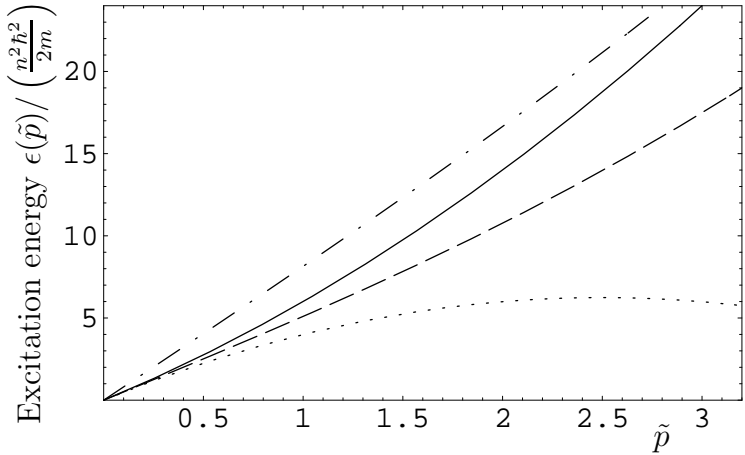}
\caption{\label{f:spectrumTL} Shown are excitation energies in the
thermodynamic limit at $\gamma = 16.3$. The full and dotted lines are
the excitations within the current Kohn-Sham approach of type I and
II from Eqs.~(\ref{eqn:TI}) and (\ref{eqn:TII}), respectively. The
dashed line is the result from the linear 
response of the bosonic LDA (\ref{elemex}).
The dash-dotted line is the Bogoliubov dispersion  (\ref{elemexBog})
which shows the wrong slope at the origin because we are out of the
perturbative regime.
}
\end{figure}

We can recover the well-known Bogoliubov
approximation in the limit of small $\gamma$ by expanding $f(\gamma) =
2\gamma^{-1} + {\cal{O}}(\gamma^{-1/2})$ to find the
usual dispersion relation
\begin{equation} \label{elemexBog}
  (\hbar \omega)_{\text{Bog}} =  \frac{\hbar^2}{2 m} n^2 \sqrt{\tilde{p}^2
     (\tilde{p}^2 +4\gamma)} .
\end{equation}
The speed of sound becomes
\[
  v_s^{\text{Bog}} = \frac{\hbar}{m} n \sqrt{\gamma} = \frac{\hbar}{m}
     \sqrt{\frac{2 n}{|a_{\text{1D}}|}}.
\]
Figure \ref{f:spectrumTL} shows the dispersion relations derived in this
paragraph for a particular value of the interaction parameter
$\gamma$.  It can be seen clearly that the Bogoliubov approximation
for the speed of sound is wrong, which is no surprise as $\gamma$ is
not small in this example. However, it is also interesting to see that
the dispersion 
derived from the bosonic LDA deviates significantly from the
fermionic LDA. 

\subsection{Limits of strong and weak interaction}

In the strongly-interacting limit we find $\nu=1$ and
$\tilde{N}=N$. In this case  $v_{\text{KS}} =0$ and
the Kohn-Sham equations become the equations for non-interacting
spinless fermions. Due to the Bose-Fermi mapping theorem of Girardeau
this gives the correct description of both the full excitation
spectrum as well as the (time-dependent) diagonal single-particle
densities. In particular we obtain the following equations for type I
and II excitations:
 \begin{eqnarray} \label{eqn:TTI}
   \epsilon^{\text{I}}(\tilde{p}) &=& \frac{\hbar^2}{2m} n^2 \tilde{p} 
     \left[ 2 \pi  + \tilde{p}\right] \\
\label{eqn:TTII}
   \epsilon^{\text{II}}(\tilde{p}) &=& \frac{\hbar^2}{2m} n^2 \tilde{p} 
     \left[ 2\pi  - \tilde{p}\right] 
 \end{eqnarray}
It is interesting to note that the single-mode equation (\ref{kmeq}),
which here is 
exactly the equation studied by Kolomeisky \cite{Kolomeisky2000a},
does not have the correct limit of Eq.~(\ref{eqn:TTI}). Instead we find
\begin{equation} \label{elemexTonks}
  \hbar \omega =  \frac{\hbar^2}{2 m} n^2 \sqrt{\tilde{p}^2
     (\tilde{p}^2 - 2 \pi^2)}.
\end{equation}
This limit underestimates the type I excitation energy. It has been
suggested in Ref.~\cite{Kolomeisky2000a} to link
dark solitary waves in the single-mode equation to the type II
excitations. The dispersion relation, however, gives only qualitative
agreement and underestimates the energy at large momenta. We want to
stress that our fermionic LDA, on the contrary, yields the exact
excitation spectrum in the Tonks-Girardeau limit.

In the limit of weak interactions we should obtain the single-mode
equation with 
\begin{equation} \label{vprimeSM}
  \phi(n) = \mu_{\text{LL}}(n) ,
\end{equation}
as was discussed in the context of the derivation of the Kohn-Sham
equations. It was also discussed that the LDA approach for finite
interactions followed in the main part of the paper cannot describe
the limit of the Bose condensate with a single occupied mode
correctly. Indeed it can be seen that Eq.~(\ref{eqn:TI}) does not
reduce to the Bogoliubov form (\ref{elemexBog}) for small $\gamma$,
although it has the correct low- and high-energy asymptotics. In
order to obtain this limit correctly, a finite size correction should
be added to the density functional.

\subsection{Static density profiles in the Thomas-Fermi approximation}
A consistency check for the static density profiles found from the
fermionic LDA can be derived from applying a Thomas-Fermi
approximation. The idea is to treat the Kohn-Sham model system of
Eq.~(\ref{KohnShamIndep}) as 
non-interacting  Fermions (equivalent to a Tonks gas) in
the external 
potential $v_{\text{ext}}(x) +
v_{\text{KS}}^{\text{LDA}}(x)$. Approximating now the density by the
Thomas-Fermi approximation or hydrodynamic LDA as in
Ref.~\cite{dunjko01} leads to  exactly the same
equations for the density as the Thomas-Fermi approximation for the
original system of interacting bosons discussed in
Ref.~\cite{dunjko01}.
This result is independent of the
specific choice of $\tilde{N}$ and provides a check for the
consistency of the LDA between the different model systems.

\section{Conclusions}

In this paper we have devised a variational scheme based on Kohn-Sham
DFT to calculate ground-state 
properties and time-dependent processes in the system of 1D bosons
with short-range interaction. The scheme will be particularly
useful for the regime of strong interaction where fermionization is
important and bosonic perturbation theory ceases to be applicable. In
contrast to the previously suggested nonlinear Schr{\"o}dinger
equation of Kolomeisky \cite{Kolomeisky2000a}, our theory has the correct
strong-interaction limit for the density and for ground-state and
excitation energies.
The degree of fermionization in the model is determined by requiring
consistency of the low-energy, low-momentum excitation
spectrum. Employing the local-density approximation yields a
parameter-free model.
Interesting applications and test cases may arise in
situation where coherence properties and long-wavelength excitations
are important. Due to the construction of our model, it is best suited
for situations with a large number of bosons and almost homogeneous
densities. A specific example may be the study of shock waves in the
homogeneous gas as studied by Damski
\cite{damski03ep:1,damski03ep:2}. In the decay of a shock wave, the usual
bosonic LDA fails due to the fact that coherence is
overestimated. Exact results are only available for the
Tonks-Girardeau limit so far. For finite interaction strengths, our
fermionic LDA
can make predictions about the importance of coherences and numerical
studies of this problem are under way. 
Another situation where the current theory can be applied
is the 1D Bose gas in a weak 
optical lattice. Interesting questions to study are the nature of
excitations and the phase diagram as well as a comparison to the
predictions of bosonic LDA.

The DFT scheme studied in this paper is based on the careful choice
of the kinetic energy functional determined by the Kohn-Sham
auxiliary independent-particle system and the rest of the functional
was approximated in the ALDA. The next logical step to improve performance of
our model is to go beyond the ALDA and employ standard correction
schemes based on current-density functional theory \cite{vignale96,vignale97}.

\ack I would like to thank William Reinhardt, Kiril Tsemekhman, and
Subhasis Sinha for encouragement and useful discussions.  Support from
the NSF, the Alexander von Humboldt Foundation,
and ECT$^*$ (Trento) is gratefully acknowledged.

\appendix
\section{Linear response equations}
Following the standard scheme of linear response theory \cite{thouless72}
we introduce a time-dependent perturbation
\[
  v_{\text{ext}}(x,t) = v_{\text{ext}}(x) + \lambda [ f^+(x)
  e^{-i\omega t} + f^-(x) e^{-i\omega t}] 
\]
in  Eq.~(\ref{KohnSham}) where
$\lambda$ is small. We look for solutions of Eq.~(\ref{KohnSham})
through first order in $\lambda$. To zeroth order we obtain the stationary
equation (\ref{KohnShamIndep}). Up to first order we expect to find
solutions of the form
\be\label{eqn:lransatz}
  \varphi_h(x,t) = \left\{ \varphi_h(x) + \lambda \sum_{p \not \in}
   (X_{hp} e^{-i \omega t} + Y^*_{hp} e^{i \omega t})
   \varphi_p(x) \right\} e^{\frac{-i\epsilon_h t}{\hbar}}
\ee
Here, the linear response of the occupied orbitals can be expanded in
terms of unoccupied ($\not \in$) orbitals because the time-evolution
according to the Kohn-Sham equations preserves the
orthonormality of the occupied orbitals to every order in
$\lambda$. Substituting the expression (\ref{eqn:lransatz}) into the
time-dependent equation (\ref{KohnSham}) and equating the first order
terms with different time-dependence separately and finally setting
$f^\pm =0$ yields the following equations for small-amplitude free
oscillations around the stationary state:
\begin{eqnarray} \label{EinsteinBog}
   \sum_{h'\in, p'\not \in}({\mathcal L}_{hph'p'} X_{h'p'} + {\mathcal
     M}_{hph'p'} Y_{h'p'}) &=& 
     \hbar \omega  X_{hp} \\ \label{EinsteinBog2}
   \sum_{h'\in, p'\not \in}({\mathcal L}^*_{hph'p'} Y_{h'p'} +
     {\mathcal M}^*_{hph'p'} X_{h'p'}) &=& 
     - \hbar \omega  Y_{hp} 
\end{eqnarray}
In the basis of stationary Kohn-Sham orbitals [i.e.\ solutions of
Eq.~(\ref{KohnShamIndep})] the matrices read
\begin{eqnarray} \label{eqn:LRorbitals}
   {\mathcal L}_{hph'p'} &=& (\epsilon_p - \epsilon_h) \delta_{pp'}
     \delta_{hh'} + \nu v'_{ph'p'h} \\
   {\mathcal M}_{hph'p'} &=&  \nu v'_{pp'h'h} ,
\end{eqnarray}
where
\be \label{eqn:me}
  v'_{ijkl} = \int \varphi^*_i(x)  \varphi^*_j(x)  \frac{d
    v_{\text{KS}}}{d n} \Big|_{n(x)} \varphi_k(x) \varphi_l(x) \; dx .
\ee
A simple basis transformation may be applied to rewrite
Eqs.~(\ref{eqn:LRorbitals}) to (\ref{eqn:me}) in a different basis,
e.g.~the coordinate representation. For the specific case of the
bosonic LDA with $\tilde{N} = 1$ we find the Bogoliubov--de Gennes
equations of Ref.~\cite{oehberg02}. In the limit $\gamma \ll 1$ the
regular Bogoliubov--de Gennes equations of Gross-Pitaevskii theory are
recovered. 

For studying the homogeneous system  where $v_{\text{ext}} = 0$
and $n$ is constant we introduce the usual box quantization with 
$\phi_m(x) = e^{i k_q x} /\sqrt{L}$ and $k_q = 2 \pi q/L$. We
find
\[
  \epsilon_i = \frac{\hbar^2 k_i^2}{2 m} + v_{\text{KS}}
\]
and $v_{ijkl} = \delta_{i-k, j-l} v'_{\text{KS}}/L$ with
$v'_{\text{KS}}\equiv dv_{\text{KS}}/dn$. Due to translational
symmetry the matrix equations block and it makes sense to label the
blocks according to the momentum transfer $\hbar k_p - \hbar k_h = 2 \pi q
\hbar/L$. With defining  ${\mathcal L}^q_{p,p'} =
{\mathcal L}_{(p-q)p,(p'-q)p'}$ we obtain
\begin{eqnarray}
   {\mathcal L}^q_{p,p'} &=& (\epsilon_p - \epsilon_{p-q}) \delta_{p, p'}
     + \frac{\nu v'_{\text{KS}}}{L}\\
   {\mathcal M}^q_{p,p'} &=&  \frac{\nu v'_{\text{KS}}}{L}
\end{eqnarray}
Due to the presence of the Fermi sphere there are restrictions on the
possible values of the indices as $k_p$ always has to be outside and
$k_{p-q}$ has to be inside.

We now consider the two cases $q=1,2$ explicitely in order to
calculate the slopes of both the type I and type II excitation
branches at zero momentum. For $q=1$ corresponding to the smallest
possible momentum transfer, the matrices ${\mathcal L}$ and ${\mathcal
M}$ have one entry each and Eqs.~(\ref{EinsteinBog}) and
(\ref{EinsteinBog2}) can be recast as a 2 by 2 matrix eigenvalue
equation. This equation can be solved easily to yield the eigenvalues
$\hbar \omega_1$ and $-\hbar \omega_1$, where $(\hbar \omega_1)^2 =
\alpha^2 +2 \alpha \nu v'_{\text{KS}}/L$ with $\alpha = \hbar^2 N 2
\pi^2/(m L^2 \nu)$ and only the positive solution is physically
relevant. Using Eqs.~(\ref{eqn:vhxc}) and (\ref{eqn:mu}) we find that
$\hbar \omega_1 / p_1 = c_s$ where $p_1 = 2\pi\hbar/L$ is the momentum
transfer for this excitation and $c_s$ is the speed of sound in the
Lieb-Liniger model given by Eq.~(\ref{eqn:sound}).  We would like to
point out that this result is independent of $\nu$ and can be traced
back to the compressibility sum rule which is granted by the LDA.

We are now going to check this result for the speed of sound by
calculating the energies for the next larger momentum transfer with
$q=2$ where 2 particle-hole excitations give the first
possibility to have different slopes for the type I and type II
excitation branches. If both branches have the same slope and yield
the same speed of sound, the energy difference between these two
excitations should vanish in the thermodynamic limit. We have solved
the eigenvalue equations (\ref{EinsteinBog}) and
(\ref{EinsteinBog2}) which are 4 by 4 with a specific symmetry.
We omit the lengthy expressions but note the
following observations: 
One of the solutions is independent of $\nu$ and gives the speed of
sound 
as in Eq.~(\ref{eqn:sound}) up to order $1/L$. The other solution does
depend on $\nu$ and for the difference between the squared slopes we find
\be
   \left[(\hbar \omega_2^{(1)})^2 - (\hbar
   \omega_2^{(2)})^2\right]/p_2^2 =   \frac{n}{2 m} v'_{\text{KS}} +
   \mathcal{O}(1/L) .
\ee
We see that this term remains finite in the thermodynamic limit unless
$v'_{\text{KS}}$ vanishes, 
which gives a stringent criterion for the choice of $\nu= N/\tilde{N}$
as discussed in Section \ref{sec:linearresponse}.
If the criterion of vanishing $v'_{\text{KS}}$ is violated, the
model will not only develop phonon branches with different speed of
sound but additionally mean-field instabilities may occur as indicated
by complex eigenvalues of the linear response equations. In the
thermodynamic limit of the
homogeneous case this occurs whenever
$\tilde{N}$ is larger than the value determined by Eq.~(\ref{eqn:nu_opt}).

% Create the reference section using BibTeX:
%\bibliographystyle{apsrev}
\section*{References}

%\bibliographystyle{prsty}
%\bibliographystyle{unsrt}
%\bibliography{/home/joachim/tex/bec,/home/joachim/tex/refs_bec,$HOME/tex/refs}

\begin{thebibliography}{10}

\bibitem{hohenberg64}
P. Hohenberg and W. Kohn, Phys. Rev. {\bf 136},  B864  (1964).

\bibitem{kohn65}
W. Kohn and L.~J. Sham, Phys. Rev. {\bf 140},  A1133  (1965).

\bibitem{runge84}
E. Runge and E.~K.~U. Gross, Phys. Rev. Lett. {\bf 52},  997  (1984).

\bibitem{vignale96}
G. Vignale and W. Kohn, Phys. Rev. Lett. {\bf 77},  2037  (1996).

\bibitem{vignale97}
G. Vignale, C.~A. Ullrich, and S. Conti, Phys. Rev. Lett. {\bf 79},  4878
  (1997).

\bibitem{girardeau60}
M.~D. Girardeau, J. Math. Phys. {\bf 1},  516  (1960).

\bibitem{cheon99}
T. Cheon and T. Shigehara, Phys. Rev. Lett. {\bf 82},  2536  (1999).

\bibitem{lieb63:1}
E.~H. Lieb and W. Liniger, Phys. Rev. {\bf 130},  1605  (1963).

\bibitem{lieb63:2}
E.~H. Lieb, Phys. Rev. {\bf 130},  1616  (1963).

\bibitem{korepin93}
V.~E. Korepin, N.~M. Bogoliubov, and A.~G. Izergin, {\em Quantum Inverse
  Scattering Method and Correlation Functions} (University Press, Cambridge,
  1993).

\bibitem{Olshanii1998a}
M. Olshanii, Phys. Rev. Lett. {\bf 81},  938  (1998).

\bibitem{petrov00}
D. Petrov, G. Shlyapnikov, and J. Walraven, Phys. Rev. Lett. {\bf 85},  3745
  (2000).

\bibitem{haldane88}
F.~D.~M. Haldane, Phys. Rev. Lett. {\bf 60},  635  (1988).

\bibitem{sutherland71}
B. Sutherland, Phys. Rev. A {\bf 4},  2019  (1971).

\bibitem{griffin95}
A. Griffin, Can. Journ. Phys. (Brockhouse issue) {\bf 73},  755  (1995).

\bibitem{nunes99}
G.~S. Nunes, Journal of Physics B: Atomic, Molecular and Optical Physics {\bf
  32},  4293  (1999).

\bibitem{Kolomeisky2000a}
E.~B. Kolomeisky, T.~J. Newman, J.~P. Straley, and X. Qi, Phys. Rev. Lett. {\bf
  85},  1146  (2000).

\bibitem{oehberg02}
P. {\"O}hberg and L. Santos, Phys. Rev. Lett. {\bf 89},  240402  (2002).

\bibitem{kim03}
Y.~E. Kim and A.~L. Zubarev, Phys. Rev. A {\bf 67},  015602  (2003).

\bibitem{chiofalo98}
M.~L. Chiofalo, A. Minguzzi, and M.~P. Tosi, Physica B {\bf 254},  188  (1998).

\bibitem{chiofalo01}
M.~L. Chiofalo and M.~P. Tosi, Europhys. Lett. {\bf 53},  162  (2001).

\bibitem{dunjko01}
V. Dunjko, V. Lorent, and M. Olshanii, Phys. Rev. Lett. {\bf 86},  5413
  (2001).

\bibitem{menotti02a}
C. Menotti and S. Stringari, Phys. Rev. A {\bf 66},  043610  (2002).

\bibitem{fetter01}
A.~L. Fetter and A.~A. Svidzinsky, J. Phys.: Condens. Matter {\bf 13},  R135
  (2001).

\bibitem{lenard64}
A. Lenard, J. Math. Phys. {\bf 7},  930  (1964).

\bibitem{girardeau00a}
M.~D. Girardeau and E.~M. Wright, Phys. Rev. Lett. {\bf 84},  5239  (2000).

\bibitem{damski03ep:1}
B. Damski, Shock waves in ultracold Fermi (Tonks) gases, preprint
  cond-mat/0306394, 2003.

\bibitem{popov83}
V.~N. Popov, {\em Functional Integrals in Quantum Field Theory and Statistical
  Physics} (Reidel, Dordrecht, 1983).

\bibitem{kulish76}
P.~P. Kuhlish, S.~V. Manakov, and L.~D. Fadeev, Theor. Math. Phy. {\bf 28},  38
   (1976).

\bibitem{jackson02}
A.~D. Jackson and G.~M. Kavoulakis, Phys. Rev. Lett. {\bf 89},  070403  (2002).

\bibitem{komineas02}
S. Komineas and N. Papanicolaou, Phys. Rev. Lett. {\bf 89},  070402  (2002).

\bibitem{levy84}
M. Levy, J.~P. Perdew, and V. Sahni, Phys. Rev. A {\bf 30},  2745  (1984).

\bibitem{thouless72}
D.~J. Thouless, {\em The Quantum Mechanics of Many-Body Systems} (Academic
  Press, New York, 1972).

\bibitem{girardeau01a}
M.~D. Girardeau and E.~M. Wright, Phys. Rev. Lett. {\bf 87},  210401  (2001).

\bibitem{schwartz77}
M. Schwartz, Phys. Rev. B {\bf 15},  1399  (1977).

\bibitem{haldane81}
F.~D.~M. Haldane, Phys. Rev. Lett. {\bf 47},  1840  (1981).

\bibitem{yang69}
C.~N. Yang and C.~P. Yang, J. Math. Phys. {\bf 10},  1115  (1969).

\bibitem{damski03ep:2}
B. Damski, Formation of shock waves in a Bose-Einstein condensate, preprint
  cond-mat/0309421, 2003.

\end{thebibliography}

%\end{document}

\end{document}